\documentclass{emulateapj}
\usepackage{ulem}
\newcommand{\ee}[1]{\mbox{${} \times 10^{#1}$}}
\newcommand{\mw}{\mbox{$ {\rm [\langle O/H \rangle_{MW}]} $}}
\newcommand{\vw}{\mbox{$ {\rm [\langle O/H \rangle_{VW}]} $}}

\newcommand{\HI}{\ion{H}{1}} 
\newcommand{\CIV}{\ion{C}{4}}
\newcommand{\CII}{\ion{C}{2}}
\newcommand{\OVI}{\ion{O}{6}}
\newcommand{\OVIa}{\ion{O}{6}a}
\newcommand{\OVIb}{\ion{O}{6}b}
\newcommand{\OV}{\ion{O}{5}}

\newcommand{\kms}{{\rm km}\,{\rm s}^{-1}}
\newcommand{\lya}{Ly$\alpha$}
\newcommand{\lyb}{Ly$\beta$}

\begin{document}

\title{A SEARCH FOR OXYGEN IN THE LOW-DENSITY LYMAN-$\alpha$ FOREST USING THE SLOAN DIGITAL SKY 
SURVEY}
\shorttitle{Oxygen in the  Ly-$\alpha$ Forest Using SDSS}

\author{Matthew M. Pieri\altaffilmark{1}, Stephan Frank\altaffilmark{1,2}, Smita Mathur\altaffilmark{1}, David H. 
Weinberg\altaffilmark{1},\\ Donald G. York\altaffilmark{3,4}, and Benjamin D. Oppenheimer \altaffilmark{5}}
\shortauthors{Pieri et al.}
\altaffiltext{1}{Department of Astronomy, The Ohio State University, 140 West 18th Avenue, Columbus, OH 43210, USA; mpieri@astronomy.ohio-state.edu}
\altaffiltext{2}{Observatoire Astronomique de Marseille-Provence, P\^ole de l'\'Etoile Site de Ch\^teau-Gombert, 38, rue Fr\'ed\'eric Joliot-Curie
13388 Marseille cedex 13, France}
\altaffiltext{3}{Department of Astronomy and Astrophysics, University of Chicago, Chicago, IL 60637, USA}
\altaffiltext{4}{Enrico Fermi Institute, University of Chicago, Chicago, IL 60637, USA}
\altaffiltext{5}{Astronomy Department, University of Arizona, Tucson, AZ 85721, USA}

\begin{abstract}

We use 2167 Sloan Digital Sky Survey (SDSS) quasar spectra to search for low-density oxygen in the Intergalactic Medium. Oxygen absorption is detected on a pixel-by-pixel basis by its correlation with \lya\ forest absorption.
We have developed a novel Locally Calibrated Pixel (LCP) search method that uses adjacent regions of the spectrum to calibrate interlopers and spectral artifacts, which would otherwise limit the measurement of  \OVI\ absorption. 
Despite the challenges presented by searching for weak \OVI\ within the Lyman-$\alpha$ forest in spectra of moderate resolution and signal-to-noise, we find a highly significant detection of absorption by oxygen at $2.7<z<3.2$ (the null hypothesis has a $\chi^2=80$ for 9 data points).

We interpret our results using synthetic spectra generated from a lognormal density field assuming a mixed quasar-galaxy photoionizing background \citep{hm01} and that it dominates the ionization fraction of detected \OVI. The LCP search data can be fit by a constant metallicity model with $\rm{[O/H]}=-2.15_{-0.09}^{+0.07}$, but also by models in which low-density regions are unenriched and higher density regions have a higher metallicity. The density-dependent enrichment model by \citet{a08} is also an acceptable fit. All our successful models have similar mass-weighted oxygen abundance, corresponding to $\mw=-2.45\pm0.06$.
This result can be used to find the cosmic oxygen density in the \lya\ forest, $\Omega_{\rm Oxy, IGM}=1.4(\pm0.2)\times 10^{-6}\approx 3\ee{-4} \Omega_{b}$. This is the tightest constraint on the mass-weighted mean oxygen abundance and the cosmic oxygen density in the \lya\ forest to date and indicates that it contains $\approx 16\%$ of the total expected metal production by star formation  up to $z=3$ \citep{b07}.

\end{abstract}

\keywords{galaxies: formation --- intergalactic medium --- quasars: absorption lines}

\section{INTRODUCTION}

The Intergalactic Medium (IGM) is the spatially undulating reservoir of baryons from which all galaxies form. During the dark ages, before formation of the first stars, this reservoir was composed of elements in proportions arising from Big Bang Nucleosythesis: hydrogen, helium and trace amounts of lithium and beryllium. Metals have been produced since the formation of the first stars and some of these metals were carried into the IGM. They are observed in the Lyman-$\alpha$ forest seen in quasar (QSO) absorption spectra at redshifts $z \lesssim 6$ (e.g. \citet{my87,c95,sc96,e00,s03,psa06}). Progress has also been made reproducing this enrichment in models of galaxy formation with extragalactic winds (e.g. \citet{mf99,mfr01,od06,pmg07,pm07}). 

Much has been learned about the degree, extent and evolution of metal enrichment of the IGM, but questions still remain unanswered. What is the precise level of enrichment in the forest, and is it consistent with that seen at the current epoch? Are underdense systems enriched? Are all enriched regions spatially associated with galaxies? In this paper, we attempt to place tighter limits on the level of enrichment in the IGM as a whole, more specifically the low-density regime ($1\lesssim \rho/{\bar \rho}\lesssim 10$), using QSO spectra from the Sloan Digital Sky Survey (SDSS, \citealt{y00}).
We have searched for \OVIa\ ($\lambda$1032\AA) at redshifts $2.7<z<3.2$. At the densities analyzed, \OVIa\ is expected to be the strongest heavy element transition in the optical band \citep{rhs97,h98}. However, the line happens to reside in the same wavelength range as the Lyman series lines at other redshifts resulting in severe confusion as to the identity of the absorption. Any search for \OVI\ must involve a sophisticated approach to overcoming this problem. 

Fitting of narrow lines or lines that also show \lya\ absorption have been performed by various authors \citep{csk02,ssr02,ssr04,bh05} in order to find individual \OVI\ systems, where high resolution spectra are available.  In contrast, pixel correlation searches (e.g. \citealt{cs98, d98, a02, s03, ph04}) are directed toward finding metals distributed throughout the forest without identifying individual systems. In the case of 
the search for \OVI, they are limited by cosmic variance in the contaminating Lyman lines and noise. The  
standard pixel (SP) correlation search (sometimes known as `pixel optical depth') for \OVI\ (e.g. \citealt{a02,ph04,a08}, hereafter A08) is not, however, a suitable approach for the detection of metals 
in the SDSS \lya\ forest because of the moderate resolution ($R=\lambda/\Delta\lambda=1800$) and signal-to-noise ratio. 
A pixel correlation search is possible in principle, but modifications to the approach are required. We have developed a tailored pixel method, which we refer to as the Locally Calibrated Pixel (LCP) search method. 
Our statistical analysis of \OVI\ absorption in the low density IGM complements the direct search for strong \OVI\ absorption in the same spectra \citep{f09a,f09b}.

While interpretation of pixel searches using high resolution, high S/N spectra is more straightforward, the small amount of high quality data leaves the conclusions subject to uncertainty owing to cosmic variance. The use of the large number of SDSS spectra ($>2000$ used here) with the LCP search method offers a large reduction in this variance (and potentially an increase in overall S/N).

All pixel correlation searches require simulations in order to ascertain the gas mass densities of enriched systems, the number densities of ions required, the ionization fraction (and so metallicity of those systems), and the impact of observational limitations and errors. This is particularly important here as the resolution of the SDSS spectra introduces a significant uncertainty in the gas mass density of enriched systems. To address this issue and interpret our search results, we have simulated each SDSS spectrum used 10 times for various models of oxygen enrichment, totaling more than 4.5 million synthetic SDSS spectra.

Noise levels in individual SDSS spectra are higher than those of high resolution spectra used in previous studies, and with this (along with the 
lower resolution) comes more uncertainty in the continuum fitting. This problem becomes increasingly severe at the 
blue end of the spectrograph, where most of the signal of \OVI\ lies. The correlation search technique 
that we outline below is calibrated to the flux level in the local spectrum to take into account variations in the continuum (hence the name `locally calibrated pixel search').

This paper is set out as follows: in \S \ref{sample} we describe the data set used, in \S \ref{method} we describe the LCP search method and results, in \S \ref{interpretation} we describe our interpretation of these results using simulated spectra, and in \S \ref{measure} we describe our measure of mass-weighted mean oxygen abundance and the cosmic oxygen density. This is followed by a discussion of the ionization corrections and other searches and, finally, the conclusions.

Following standard notation, we write $\rm{[O/H]=log(O/H)-log(O/H)_\odot}$, where O/H is the ratio of oxygen to hydrogen (by number). We adopt $\rm{ log (O/H)_\odot=-3.13}$ estimated for the solar envelope by \citet{gns96}, for ease of comparison with other studies of oxygen enrichment of the IGM. While some analyses based on more detailed stellar atmosphere models have argued for lower (O/H)$_\odot$ \citep{ags05,c08}, stellar interior models and helioseismology strongly support the Grevesse et al. abundance scale \citep{ba04,dp06}.  For values of solar abundance taken from \citet{ags05} and \citet{c08}, scale our [O/H] results by $+0.21$ and $+0.11$ respectively.

\section{THE SAMPLE OF SPECTRA}
\label{sample}

The \OVI\ doublet starts to become visible at the blue end of SDSS quasar spectra at redshifts of $z_{\rm abs} \geq 2.7$. Hence, we retrieved QSO spectra from the SDSS Third Data Release (DR3; \citealt{aetal05}) beyond this lower boundary redshift using the QSO Absorption Line Systems (QSOALS) project database \citep{y06}.

The spectra we used were fitted with an automatic continuum procedure developed by Arlin Crotts and used in the QSO absorption line data base of the SDSS QSOs \citep{y06}. The continuum is derived by discarding absorption-contaminated pixels until the variance in the remaining pixels corresponds to that prescribed by the noise model for the measurement.  This provides 
a smoothed estimate of the spectrum, forming a ``continuum" for the measurement of narrow lines over a range of pixels that varies with the large scale variations in the SDSS spectrum. At a minimum (e.g. near narrow emission lies from the QSO or the sky, or at the steep drop at the edge of a broad absorption line) the smoothing is over a 20 pixel region (between 18\AA\ to 24\AA\ in the forest used here). In smooth regions of the spectrum, outside the Lyman-$\alpha$ forest, the method produces continua accurate to about 2\% when the photon noise is small enough to permit such precision.
The method produces a reasonable estimate for the continuum in the Lyman-$\alpha$ forest for our purposes. Note that the rare cases of a Lyman-$\alpha$ BAL or the more common cases of an O VI BAL will be fitted away and this is a desirable feature for our purposes.

There are rarely rapid continuum changes to be expected over 30 \AA{}, and certainly no significant non-linear trends. This gives us a good enough continuum for application of
 a locally calibrated continuum for the \OVI\ absorption measurement, as described in \S\ref{method}.  The continuum fitting method may subtract large-scale structures in \lya\ absorption on scales $>2800 ~\kms$, but this effect will be minimal as indicated by the good agreement between the probability distribution function (PDF) of the \lya\ optical depth in the data and the models as shown in Figure \ref{pdfcompare} below. This continuum fitting method results in a  $\sim16\%$ systematic underestimate of the continuum (and so an overestimate of the transmitted flux), measured as an equivalent underestimate of the mean flux decrement in the sample with respect to values in the literature (e.g \citealt{m00, k07}). We have taken this into account by introducing the same systematic to our simulations as described in \S\ref{synthetic spectra}.

We do not exclude BAL quasars from the sample used in this study. We have tested the impact of these BAL QSOs by performing the LCP search excluding the 430 QSOs that are identified by \cite{g09} as BAL QSOs and fall into our sample. The impact of this modification is much smaller that the $1\sigma$ error bars.

As can be seen in  Figure 3 of \citet{f09a},
the majority of the sources have $i${} band magnitudes between 19.5 and 20.5. The average signal-to-noise ratio of the spectra within the region of interest for us is only about 2.5, much less than the overall signal-to-noise ratio of the spectrum, since the flux levels within the Lyman-$\alpha${} forest are much lower than those beyond it. This figure also shows the distribution of these average signal-to-noise ratio values with the emission redshifts of the sample sources. Particularly towards higher redshifts, it is apparent that the increasing density of the forest and the general faintness of the sources lead to a severe decrease in the signal-to-noise ratio.

In order to ensure that the noise characteristics of the sample can be adequately reproduced in the simulated spectra, we have performed an analysis of the behaviour of the noise in the full data set at every observed wavelength. The uncertainty in the flux measurement, $\sigma (f_{\lambda},\lambda)$, is the sum in quadrature of two terms,
 the readout noise of the detector system and the photon noise. The former is a Gaussian distribution with $\sigma_{ro}(\lambda)=k_1$, where $k_1$ is a constant at a particular wavelength, independent of the incident flux. The latter is a Poisson distribution with $\sigma_{ph}(f_{\lambda},\lambda) = k_2 \times \sqrt{f_{\lambda}}$, where $k_2$ is a constant at a particular wavelength.
 
\begin{figure}
\includegraphics[angle=270,width=\columnwidth]{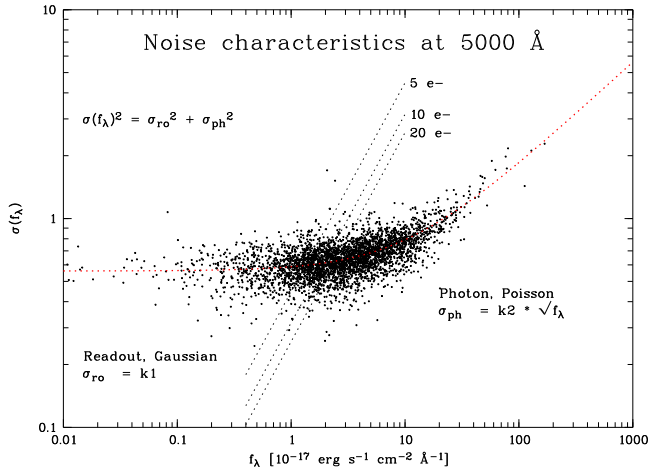}\label{sigma_5000A}\caption{Example of the noise characteristics of the sample spectra at 5000 \AA. Plotted in black are the estimated noise levels in real flux, $\sigma(f_{\lambda})$, versus the flux at 5000 \AA. The noise can be separated into two components, the detector dependent read-out noise, $\sigma_{\rm ro}$, and the flux dependent photon noise, $\sigma_{\rm ph}$. The dotted red line shows the two parameter fit to the data, as described in the text. The dashed black lines indicate the location of the noise if it were purely dominated by the Poisson statistics of the photon noise with 5, 10 and 20 counts. Where the photon noise dominates, the counts are clearly above 20, and so this Poisson distribution is well described by a Gaussian appoximation. Hence the noise distribution is Gaussian at all levels of flux. We have confirmed this for the complete wavelength range covered by SDSS.} 
\label{noisechar}
\end{figure}

 At each wavelength, we fit the distribution of the error estimate in the flux provided by the QSOALS database with this two parameter model, and hence are able to describe the complete noise characteristics of the underlying SDSS sample by the wavelength dependent $k_1(\lambda)${} and $k_2(\lambda)$. Figure \ref{noisechar} shows an example of the noise distribution as a function of the estimated flux, $f_{\lambda}$, at a specific wavelength.  We have used this formalization of the noise characteristics to produce a realistic representation of observing conditions in synthetic spectra as described in \S \ref{synthetic spectra}. This figure also illustrates how the noise may always be characterized as a Gaussian distribution, since the (Gaussian) read out noise dominates where the Poisson nature of the photon noise is significant.

\section{THE LOCALLY CALIBRATED PIXEL SEARCH}
\label{method}

Pixel correlation searches for the detection and investigation of metal enrichment in the IGM have, 
until now, been limited to high resolution quasar absorption spectra. In these spectra, the 
optical depth of every pixel of the Lyman-$\alpha$ forest can be characterized as a measure of the 
density in the IGM. Where the \lya\ absorption is saturated (and so the measured \lya\ optical depth measure is noisy), higher order Lyman lines can be used to reconstruct the \lya\ optical depth.

Ionization species other than neutral hydrogen can produce their own 
forest of absorption. In particular there is a forest of five-times-ionized oxygen (\OVI) and a forest 
of three-times-ionized carbon (\CIV). These metal line forests can also be dealt with on a pixel-by-pixel basis at each redshift. Hence, one can collect \lya\ and \OVI\ or \CIV\ pixel pairs, bin by \lya\ 
optical depth, and one would see a clear trend in metal line optical depth (using only basic assumptions of
ionization and metallicity, which we will return to throughout this paper).

Difficulties arise when the measured optical depth of the metal pixels is not entirely due to the 
intended species. Various techniques have been used in the SP search to minimize the degree of contaminating absorption, 
but it cannot be removed entirely, and it tends to dominate where absorption is weakest. These techniques involve correction of contaminating Lyman series lines in the \OVI\ region of the spectra by removing expected higher order lines (based on the \lya\ absorption at the same redshift and longer wavelength), or the use of both \OVI\ lines to derive a minimally contaminated signal. 

As is done in all pixel correlation searches in quasar spectra, we compared the absorption from a species at a fixed 
redshift with the absorption from another species at the same redshift. Where the absorption is correlated, a detection is found.  In this analysis we search for \OVIa\  ($\lambda 1032$\AA) absorption correlated with Lyman-$\alpha$ ($\lambda1216$\AA) absorption and make no use of the \OVI\ doublet line (\OVIb) at $\lambda1038$\AA. We produce these pixel pairs for a range of redshifts and on a 
grid set by the spectral binning of the \OVIa\ pixels (the \lya\ absorption is interpolated). Nearby pixels with a wavelength close to the \OVIa\ wavelength are used to characterize continuum errors and contaminating absorption. This procedure is explained in more detail below.

We have simplified the algorithm in recognition of the fact that there are features of the SP search that 
are unsuitable in the context of our local calibration method. These techniques are directed toward subtracting contamination. We do not seek to remove contaminating absorption; as described above, we {\it characterize} it, and, as shall be seen below, the LCP analysis does this automatically.

The lower redshift limit for every quasar in the sample is set by the blue end of the spectrograph (around $3800$\AA) or the Lyman-$\gamma$ line at the quasar redshift, whichever results in a 
higher minimum redshift. 
We discard pixels with Lyman-$\gamma$ as they provide little extra information but modify the 
distribution of absorbers in \OVI\ signal enough to introduce unnecessary uncertainty. 
Note that this requirement also means that no absorption at the Lyman limit is possible. 
We rule out regions within 
$5000 ~\kms$ of the QSO emission redshift to eliminate most effects due to the QSO and its environment \citep{w08}.  This sets the upper redshift limit and 
also results in an \OVI\ signal entirely within the Lyman-$\beta$ forest (since the separation between \OVIa\ and \lyb\ absorption at the emission redshift is $1,804 ~\kms$ in the \OVI\ frame).

\begin{figure}
\centering
\mbox{
\includegraphics[angle=0,width=.98\columnwidth]{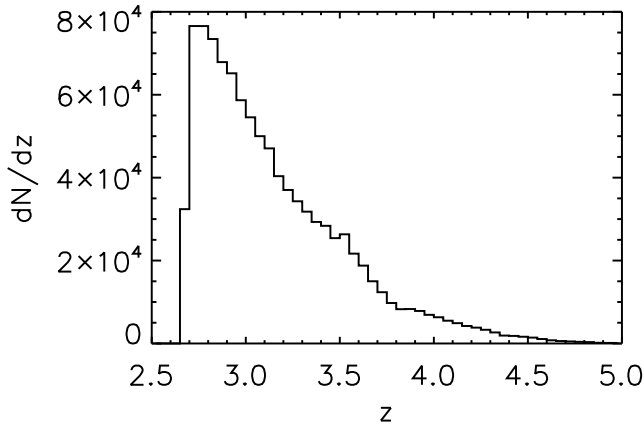}}
\caption{Redshift distribution of pixel pairs available for the analysis in the SDSS sample. The lower envelope is set by the lowest redshift systems observable in \OVI\ at the blue end of the spectrograph. The upper envelope is set by $5000 ~\kms$ below the emission redshift of QSOs in the sample}
\label{dndz}
\end{figure}

We combine the pixel pairs for each QSO into one combined sample for all the SDSS spectra in the required redshift 
range producing over a million pixel pairs. The redshift distribution of these data is shown in Figure \ref{dndz}.

\subsection{\HI\ Absorption}
\label{hi}

The determination of the degree of \HI\ absorption is largely unchanged from the SP search. The optical 
depth to Lyman-$\alpha$ absorption $\tau_{\rm Ly\alpha}$ is determined for each pixel. We make the same noise 
requirement as other searches: the flux transmission must be $\sigma_{\rm noise}/2$ from either the zero level or the continuum, where $\sigma_{\rm noise}$ is the continuum scaled error estimation in the continuum fitted spectrum. This is a 
rather strict criterion in the context of the SDSS sample and discards $\approx20\%$ of pixel pairs.
We do not attempt to extend the analysis to the higher density regime, where \lya\ absorption is saturated, by using higher order Lyman lines to derive the \lya\ optical depth. This is a regime to which other methods are better suited \citep{f09a,f09b}. It should also be noted we assume that all \OVI\ is associated with \HI\ absorption, but under most physical conditions this is a fair assumption.

The large amount of data at our disposal allows us to aggressively discard less reliable measures of $\tau_{\rm Ly\alpha}$. Hence we require that $\sigma_{\rm noise}<0.1$, which discards a further 70\% of the pixel pairs, so the feature we are searching {\it by} (\lya) is as clean a signal as we can make it while preserving a large sample of data. As we shall show, the feature we are searching {\it for} (\OVIa) is treated with no such noise requirements,  and brute force averaging is used to provide a clean signal.

\subsection{Measuring the Bulk \OVI\ Absorption}
\label{ovi}

Any measure of weak correlated \OVI\ absorption must deal with the biggest uncertainty: the continuum level. The 
primary goal for the following technique is to remove the impact of errors in the continuum fitting from the derived 
signal. Rather than simply taking the transmitted flux at the position in the spectra corresponding to the required \OVI\  absorption we determine the ratio of the transmitted flux at this point compared to that of a neighboring pixel. Where this pixel is 
sufficiently close, this gives us a measure of the absorption with the local error in the continuum fitting subtracted. It 
also results in a subtraction of average levels of contaminating \lya, \lyb\ and \OVIb\ absorption along with a measure 
of nearby \OVIa\ absorption. The ratio of transmitted flux, F, at the \OVIa\ pixel to that on a locally calibrating pixel is
\begin{equation}
\frac {F_{\rm OVIa}} {F_{l}}=\frac{f_c \exp[-\tau_{\rm OVIa}]\exp[-\tau_{u}] +{\rm noise}}
 {f_{c,l}  \exp[-\tau_{{\rm OVIa},l}] \exp[-\tau_{u, l}] +{\rm noise}_l},
 \label{fovi/fl}
\end{equation}
where $f_c$ is the flux level at the 
continuum, $\tau_{\rm OVIa}$ is the optical depth to \OVIa\ absorption, `noise' is the Gaussian instrumental noise, 
the subscript `$l$' indicates a local value and the subscript `$u$' refers to uncorrelated contaminating absorption. This 
uncorrelated absorption is dominated by \lya\  and \lyb\ absorption, but it also includes contamination by the second 
member of the \OVI\ doublet ($\tau_{\rm u}=\tau_{\rm Ly\alpha}+\tau_{Ly\beta}+\tau_{\rm OVIb}$), all at different redshifts. $\tau_{\rm OVIa,l}$ is the local \OVIa\ absorption, which is also uncorrelated where the appropriate local pixels are used, as we shall argue in \S \ref{locpix}.

Multiple local pixels are used, and each one provides a new measure of spectral 
characteristics local to our required \OVIa\ pixel. Operationally, we treat each local pixel as providing a new measure of $F_{\rm OVIa}/F_l$
and so a new pixel pair of $\tau_{\rm Ly\alpha}-F_{\rm OVIa}/F_l$. In this way, the final statistical error is dominated by the number of \lya\ - \OVI\ pixels and not by the noise in the local pixels.

\subsection{The Aggregation of Pixel Pairs}

Since we have a large sample of spectra with comparable characteristics and a limited redshift range we aggregate all 
acceptable pixel pairs from all our spectra. We use pixels in a redshift range of $2.7<z<3.2$ both to keep 
our data as homogeneous as possible and to limit ourselves to the subset that provides the strongest 
signal. This limits us to 2167 QSO spectra in our sample, out of the 5767 $z>2.3$ quasar spectra in SDSS DR3.

Once we have a list of  $\tau_{\rm Ly\alpha}-F_{\rm OVIa}/F_l$ pairs, as set out above, we bin them by their  $
\tau_{\rm Ly\alpha}$ and take the median of the $\tau_{\rm Ly\alpha}$ and the $F_{\rm OVIa}/F_l$  in order to 
produce our final search for \OVIa. We use the median as an outlier- and noise-resistant measure of the typical absorption.

Since $\tau_{u,l}$ and $\tau_u$ are drawn randomly from the same distribution, and (where the local pixel is sufficiently close)  $f_{c,l}\approx f_{c}$, the median$(F_{\rm OVIa}/F_l)$ is a quantity that characterizes correlated \OVIa\ absorption normalized to a factor dependent on the mean \OVIa\ optical depth.
The precise level of this normalization is unimportant in the following results; as with other pixel correlation searches, a statistically significant correlation is required for a detection of metals. It is, however, broadly indicative of the average \OVIa\ absorption level. It is also notable that this level is recovered in the null tests that follow.

\subsection{Choice of Local Pixels}
\label{locpix}

For local pixels we chose those that are i) sufficiently far away from the \OVIa\ search 
pixel to avoid comparison with pixels within the same extended complex associated with the same enriched region, ii) sufficiently close to avoid producing spurious correlations with the Lyman-$\beta$ and \OVIb\ absorption, and iii)
sufficiently close to measure the same continuum level. The second requirement ensures that the third requirement is satisfied.
Lyman-$\beta$ is stronger and correlated over a larger velocity range compared with \OVIb, hence we design our choice of local pixels for avoidance of this line. We also require that the choice of local pixels are mirrored on each side of the search pixel in order to cancel continuum level error trends across the line.

We take 18 local pixels in total for this 
analysis: the 9th to the 17th pixel from the search pixel at both lower and higher wavelengths. This results in separations of $\gtrsim 600 ~\kms$ and $\lesssim 1200 ~\kms$ from the \OVIa\ and \lyb\ transitions, which is approximately $9-16$\AA\ in the observed frame at $z=3$. This provides $\approx 800,000$ pixel pairs in all the searches that follow.

We tested our choices of locally calibrating pixels by treating each of them individually as a `search target' (searching for correlations with \lya) and finding the expected null result. Pixels within $600 ~\kms$ of \OVIa\ or \lyb, on the other hand, yielded signals of contaminating correlated absorption. The  $600 ~\kms$ scale is characteristic of the winds from Lyman break galaxies (e.g. \citealt{p01,a03}) and the maximum clustering scale found for \CIV\ \citep{a03,a05,psa06,s06}, so these results seem reasonable for \OVI. Correlations in \lyb\ extend over greater scales than this due to large scale structure \citep{m06}, but such lines are somewhat stronger than expected here and our test searches confirm this. In principle \CII ($\lambda1036$\AA) could produce correlated absorption within our calibrating pixel range, but in practice it appears to be too weak.

In test searches we detected the doublet line \OVIb\ at a consistent level with that of the \OVIa\ line. Since this line merely constitutes a parallel but less sensitive measure of \OVI\ that does not contribute to the search algorithm, we have not used this information.

\subsection{\OVI\ Search Results}

\begin{figure}
\centering
\mbox{
\includegraphics[angle=0,width=0.95\columnwidth]{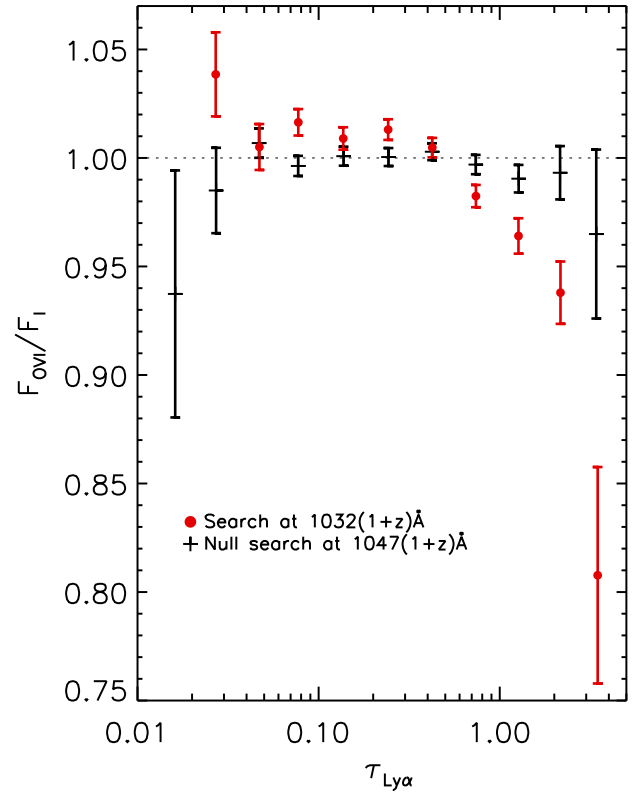}}
\caption{The LCP search for \OVIa\ in the SDSS \lya\ forest from pixels with $2.7<z<3.2$.
The red filled circles are the median $F/F_l$ on and off the 1032\AA\ \OVIa\ resonance line vs the $\tau_{\rm Ly\alpha}$ at the same redshift. Where $F_{\rm OVI}/F_l$ is below (above) 1 there is a greater (lower) than average degree of \OVI\ absorption. A correlation in this plot is indicative of a detection of \OVI.  The black `-'s are the null case of searching on and off 1047\AA.
}
\label{searchandnull}
\end{figure}

Once the pixel pairs required are combined and binned by their $\tau_{\rm Ly\alpha}$, we find their medians to plot $F_{\rm OVI}/F_l$ vs. $\tau_{\rm Ly\alpha}$  as shown in Figure \ref{searchandnull}. We required that more than 200 pixel pairs be 
available for a bin to be used. In all panels the error bars are produced by bootstrap resampling with 100 realizations and a bootstrap element of 1 SDSS QSO. The median is found for every realization, and the error bars are then taken to 
be the $1\sigma$ deviation in the  distribution of medians.

Figure  \ref{searchandnull} shows $F_{\rm OVI}/F_l$ vs. $\tau_{\rm Ly\alpha}$   using pixels in the redshift range $2.7<z<3.2$. There is a clear correlation between $\tau_{\rm Ly\alpha}$ and $F_{\rm OVI}/F_l$, and this correlation extends to all measured values of $\tau_{\rm Ly\alpha}$. There is both excess \OVI\ absorption in pixels with strong \lya\ absorption and a deficit of \OVI\ absorption ($F_{\rm OVI}/F_l>1$)  in pixels where \lya\ absorption is weak.

The null test shown in Figure \ref{searchandnull} is performed using the same technique as set out above 
with the only modification that the restframe wavelength for correlations is no longer the \OVIa\ restframe wavelength 
($\lambda1032$\AA)  but $\lambda=1047$\AA. 
As can be seen, the search approach passes the null test by recovering the expected result of  $F_{\rm 1047}\approx F_l$, where neither an excess nor a deficit in absorption is seen as no signal is present.

Figure \ref{searchandnull} is the main observational result of this paper. In high resolution spectra there is an approximate one-to-one relation between $\tau_{\rm Ly\alpha}$ and gas overdensity $\rho/\bar{\rho}$ \citep{r97,c98, w98}. At SDSS resolution this is no longer the case, so the interpretation of this result in terms of enrichment of the IGM must rely on synthetic spectra drawn from an underlying physical model.

\section{INTERPRETATION USING SYNTHETIC SPECTRA}
\label{interpretation}

\subsection{Production of Synthetic Spectra}
\label{synthetic spectra}

We use an approach for the production of simulated line-of-sight density distributions that was set out in \citet{ph04}. We restate this method here, but for a full description refer to that paper. We use cosmological parameters consistent with WMAP5 \citep{k08} ($\Omega_m=0.268$, $\Omega_\Lambda=0.732$, $\Omega_b=0. 0441$, $\sigma_8=0.776$ and $H_0=70.4~\kms {\rm Mpc^{-1}}$) in these simulations

We start with a 3D power spectrum of dark matter density fluctuations taken from \citet{ebw92}. This is normalized using the cosmological parameter, $\sigma_8$ (the amplitude of clustering on an $8h^{-1} {\rm Mpc}$ scale at the present epoch), and the linear growth of structure at the required redshift. A Jeans length filtering ($\sim0.9h^{-1}$Mpc comoving in the surveyed redshift range) to remove structure on small scales is used to describe baryonic pressure effects.

Two 1D power spectra are produced by integration of the 3D power spectrum \citep{kp91}, and Gaussian random realizations of structure are generated. These 1D realizations of structure are combined to produce coupled density contrast and velocity fields. This velocity field,  $v(x)$, is a non-dynamical approximation based on the linear density field, $\delta(x)$. For more details on this approach see \citet{bd97} and references therein.

Further following the method of \citet{bd97}, we convert our linear density field to a lognormal distribution, using the mapping
\begin{equation}
\rho/\bar{\rho}(x)={\rm exp}[\delta(x)-\langle\delta^2\rangle/2],
\end{equation}
which they find reproduces the probability distribution function of non-linear structures. It should be noted that this method fails to reproduce the clustering of non-linear structures \citep{v02}. The density field produced has a resolution of $0.01 {\rm Mpc}$ comoving, which substantially over-samples the resolution required by the SDSS sample but is adopted in order to ensure that the physics of the IGM is well described. 

The temperature of the IGM in these simulations is set by the balance between photoionization heating and adiabatic cooling and follows the power-law relation
\begin{equation}
T=T_0(\rho/\bar{\rho})^{0.4}
\label{equofstate}
\end{equation}
\citep{hg97}, where $\rho/\bar{\rho}$ is the overdensity and the temperature at mean density is $T_0=2\ee4 {\rm K}$. The temperature is required to calculate the Doppler parameter for this gas.

For a given set of cosmological parameters and a helium fraction of 0.242 by mass, it is trivial to calculate the hydrogen density. Oxygen is added in quantities as described in the following section. The ionization fractions of hydrogen and oxygen are calculated using CLOUDY version 08.00 \citep{f98} and the \citet{hm01} (hereafter HM01) quasar and galaxy UV background model (designated QG) with a 10\% escape fraction from galaxies.  While these calculations do take into account collisional ionization, the lack of heating other than via photoionization in our simulations makes the conditions for it to occur rare. As such, we predominately interpret the detectable \OVI\ to be outside of this collisionally ionized, hot gas phase. This assumption is supported by observations that the bulk of previously detected  \OVI\ lines in the \lya\ forest are photoionized \citep{csk02,b02,ssr04} based on their line widths. This approach provides fully self-consistent models since regions with $\rho/{\bar \rho}<10$ will be overwhelmingly photoionized and we find this regime dominates our measured \OVI\ (see \S\ref{comparison}).

\begin{figure}
\centering
\mbox{
\includegraphics[angle=0,width=.95\columnwidth]{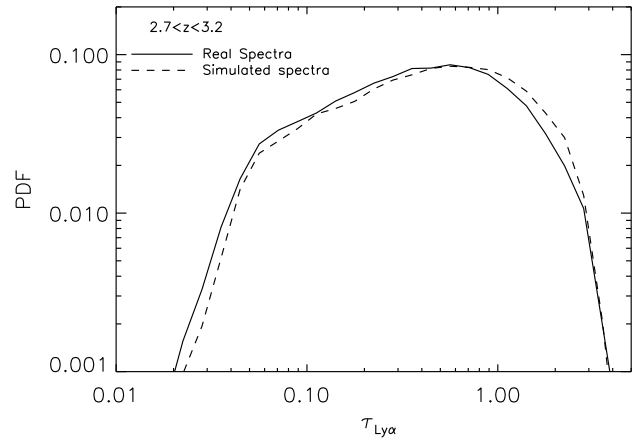}}
\caption{The PDF of \lya\ optical depths in the observed sample ({\it solid} line) and one simulated suite ({\it dashed} line). The simulated sample was produced with a quasar and galaxy HM01 UV background, the intensity of which was renormalized to match the observed distribution giving an ionization rate of $\Gamma=13\ee{-13} {\rm s^{-1}}$.}
\label{pdfcompare}
\end{figure}

We calculate the optical depths for \lya, \lyb, \OVIa\ and \OVIb\ using
\begin{equation}
\tau (v)=\frac{\pi e^2 \lambda f} {m_e c H_z}\int \frac{n(v^\prime)}{b(v^\prime)}{\rm exp}\Big[-\Big(\frac{v^\prime-v_0}{b(v^\prime)}\Big)^2\Big] dv^\prime,
\end{equation}
where the transition dependent parameters are the rest frame wavelength $\lambda$, the oscillator strength $f$, the number density of absorbing particles  $n$, and the Doppler parameter $b$. $H_z$ is the redshift dependent Hubble parameter, $v$ is the line-of-sight velocity, $v_0$ is the velocity of the Hubble flow, $e$ is the electron charge, $m_e$ is the electron mass, and $c$ is the speed of light. This is integrated over the narrow range of velocities that contribute significantly to the calculation (effectively a few times the Doppler parameter).

We add instrumental broadening to match the SDSS spectral resolution. As stated in \S\ref{sample}, we have performed an analysis of the characteristic $\sigma(f_\lambda,\lambda)$ in our sample, and we use this to add Gaussian noise to the spectra. In order to achieve this we use the continuum level in the real spectra to rescale the simulated transmission flux to a flux, $f_\lambda$.  The photon noise is well described by a Gaussian distribution where it dominates, and the readout noise is also a Gaussian, so treating the combined noise $\sigma(f_\lambda,\lambda)$ as a Gaussian is a good approximation (\S \ref{sample}). Finally, we rescale our transmitted fluxes up by 16\% in line with the systematic underestimate of the continuum in our sample (\S \ref{sample}).

The intensity of the UV background was renormalized (see \S\ref{ioncorr}) to provide a match of the PDF of optical depths derived from our synthetic spectra with that of the observed sample. We obtain agreement  with an ionization rate of $\Gamma=13\ee{-13} {\rm s^{-1}}$ as is shown in Figure \ref{pdfcompare}. The quality of this agreement is sufficiently good to populate the bins in $\tau_{\rm Ly\alpha}$ with statistics that are representative of the LCP search in the data.

\subsection{The Monte Carlo Approach}
\label{montecarlo}

For every oxygen abundance pattern we simulate a full suite of 2167 SDSS QSO spectra and do this ten times. For each one we perform an LCP search for \OVI, and the resultant $F_{\rm OVI}/F_l$ vs. $\tau_{\rm Ly\alpha}$  is averaged over the ten suites to obtain a model prediction (so that the quality of the fit between the model and observed LCP searches is dominated by the error in the observed search). We have used this set of 21,670 line-of-sight density distributions with varying levels of oxygen abundance to interpret our observations. Each of these lines-of-sight require the simulation of a 1D density distribution of length around $1000 h^{-1}{\rm Mpc}$. The production of this quantity of data is not practical with fully numerical simulation methods given current computer processor limitations. This underlines the value of the analytical technique described in the previous section. 212 models of oxygen enrichment have been tested, totaling more than 4.5 million synthetic spectra produced. 

In the following analysis we do not include the measure at $\tau_{\rm Ly\alpha} \approx 3.3$ (see Figure \ref{searchandnull}). The inclusion of this point results in poor agreement with the models. This is consistent with the findings of A08, that high  $\tau_{\rm Ly\alpha}$ absorption results from a different population of \OVI\ absorbers that are most likely collisionally ionized.

\begin{figure}
\centering
\mbox{
\includegraphics[angle=90,width=.95\columnwidth]{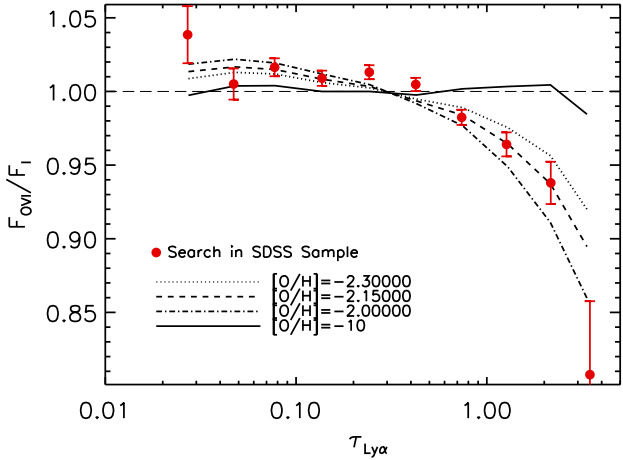}}
\caption{The LCP search compared with models of density independent [O/H]. The red filled circles are the LCP search as in Figure \ref{searchandnull}. Four models are shown: the {\it solid} line is the null case where no oxygen is added, the {\it dashed} line shows the best fitting case of a constant [O/H]$=-2.15$. A further two models show oxygen metallicities that are ruled out: [O/H]$=-2.00$ ({\it dot-dashed} line) and [O/H]$=-2.30$ ({\it dotted} line). Note that the $\tau_{\rm Ly\alpha} \approx 3.3$ point is not included in the fits. We draw detailed conclusions on metallicity constraints in \S\ref{measure}.
}
\label{plotsimav}
\end{figure}

\subsection{Comparison of Real and Synthetic Spectra}
\label{comparison}

We add oxygen with a constant [O/H] to our synthetic spectra and have varied this [O/H] to find a good fit to the observed search. Figure \ref{plotsimav} shows a comparison of the observed search (filled circles) with four models. Once more we see that our LCP search responds well to the null test (solid line) by returning a result of  $F_{\rm OVI}\approx F_l$ when no oxygen is added (the {\it reduced} $\chi^2=10$). In the best fitting case, $\rm{[O/H]}=-2.15$ (dashed line), we find good agreement with the observed result, and there is no need for the addition of more parameters in the fit. Models with constant $\rm{[O/H]}=-2.0$ (dotted line) or  $\rm{[O/H]}=-2.3$ (dot-dashed line) are clearly ruled out and indicate that the LCP search in the SDSS sample is sensitive to small changes in the abundance of oxygen in the IGM. 

\begin{figure}
\centering
\mbox{
\includegraphics[angle=90,width=.95\columnwidth]{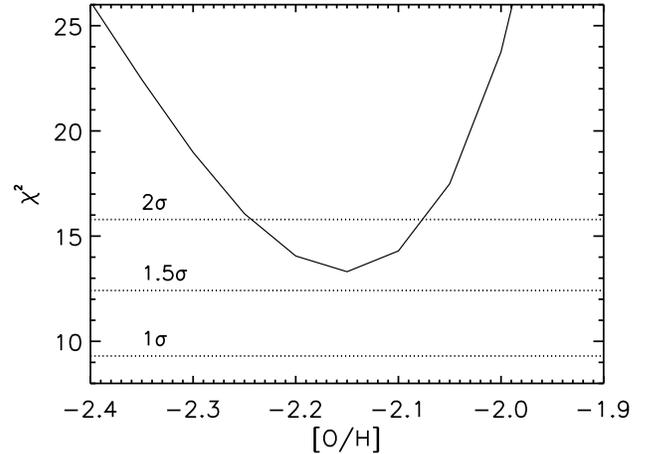}}
\caption{The $\chi^2$ test of fit ({\it solid} line) between LCP searches performed on our models with varying density independent [O/H] and our observed LCP search. The {\it dotted} lines show the $1\sigma$, $1.5\sigma$ and $2\sigma$ confidence levels indicating the range of [O/H] that constitutes a good fit.}
\label{chinocut}
\end{figure}

We can quantify the quality of the fit by using the $\chi^2$ test where the error bars are given by the boot-strapping of the observed data.  We have done this for varying [O/H] as shown in Figure \ref{chinocut} (solid line). The $1.5\sigma$ and $2\sigma$ level are shown as the dotted lines. It is clear that our best fitting model with $\rm{[O/H]}=-2.15$  constitutes an adequate fit to the observed search. We can give this measure a $2\sigma$ confidence interval of $\rm{[O/H]}=-2.15_{-0.09}^{+0.07}$. These models constitute the range of constant [O/H] models favored by the LCP search. The quality of this fit is suggestive of an underlying density dependence to the [O/H] signal (as found by A08), but the statistics do not warrant a two parameter fit.

\begin{figure}
\centering
\mbox{
\includegraphics[angle=90,width=.95\columnwidth]{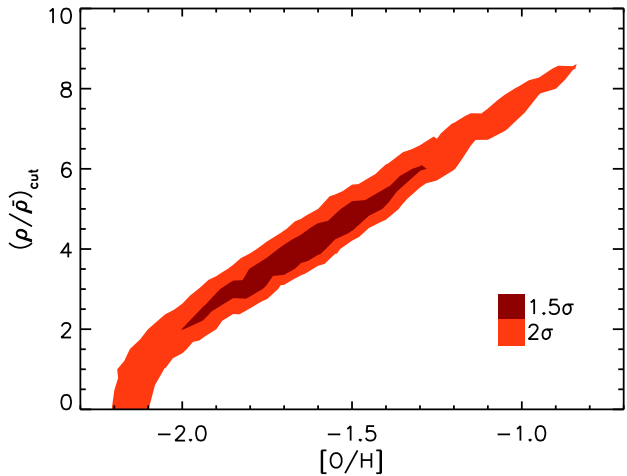}}
\caption{The $1.5\sigma$ (dark red) and $2\sigma$ (orange) likelihood of fit between LCP searches performed on our models and our observed LCP search (with two parameters and nine data points). The models have a constant [O/H] above a threshold gas overdensity, $(\rho/\bar{\rho})_{\rm cut}$. Below this threshold no oxygen is added. The contours characterize the range of densities for which this search for \OVI\ is sensitive. The likelihood is largely unaffected by the addition of a threshold at $\rho/\bar{\rho}_{\rm cut}=1$, indicating that the search is not sensitive to metals in underdense systems. Where a good fit can no longer be obtained for a particular value of $(\rho/\bar{\rho})_{\rm cut}$, this constitutes the upper limit on the density range to which we are sensitive. }
\label{chicontour}
\end{figure}

We may not be able to constrain the density dependence of this oxygen enrichment, but we can place limits on the density range in which it resides. Hence we introduce a new parameter: a gas overdensity threshold, $(\rho/{\bar \rho})_{\rm cut}$. Where $\rho/{\bar \rho}<(\rho/{\bar \rho})_{\rm cut}$, we add no oxygen, and where $\rho/{\bar \rho}>(\rho/{\bar \rho})_{\rm cut}$ we add the constant [O/H] specified. Figure \ref{chicontour} shows a likelihood plot of the fit to the SDSS data set for 208 models (with two parameters and nine data points).  Where $(\rho/{\bar \rho})_{\rm cut}=0$, no threshold is used and the models correspond to those used in Figure \ref{chinocut}. There is little or no change in the quality of the fit going from $(\rho/{\bar \rho})_{\rm cut}=0$ to $(\rho/{\bar \rho})_{\rm cut}=1$, and this indicates that our LCP search for \OVI\ in the SDSS QSO sample is not sensitive to metals in underdense regions of the Universe. 

As the threshold is raised in our models a degeneracy becomes apparent, as seen in Figure \ref{chicontour}: removing metals from low-density regions requires one to add more metals to higher density regions to compensate, and it is {\it still possible to obtain a good fit}. This indicates that our search is rather insensitive to the density of the systems enriched. As expected, the tight relationship between $\tau_{\rm Ly\alpha}$ and overdensity seen in high resolution spectra \citep{r97, c98, w98} breaks down due to the moderate resolution of the SDSS sample. Dense systems can always produce a low apparent $\tau_{\rm Ly\alpha}$ by virtue of the instrumental broadening over multiple pixels, but the key question is {\it to what extent those systems can reproduce the observed signal?} The agreement begins to break down at $(\rho/{\bar \rho})_{\rm cut}\approx6$, where we have 13\% confidence ($1.5\sigma$) in our fit to the observed result, and it breaks down completely at $(\rho/{\bar \rho})_{\rm cut}\approx8.5$, where we have a 5\% confidence ($2\sigma$) in our fit. This breakdown occurs because there are proportionately too few higher density systems to obtain a fit to the search curve even with arbitrarily high metallicity. The best fit is obtained with $(\rho/{\bar \rho})_{\rm cut}=4$, where the fit has $\chi^2=9.4$ (23\% confidence for two parameters and nine data points).

While Figure \ref{chicontour} shows that a range of two-parameter models can fit the data, we will show in \S \ref{measure} that the mass-weighted mean oxygen abundance is well constrained. To demonstrate this, we must first address the issue of scatter in the metallicity.

\subsection{Scatter in the Metallicity}
\label{scatter}

\citet{s03} (hereafter S03) and A08 found that a lognormal scatter in the metallicity is a good fit to all percentiles in their search. We assume the same scatter  in order to have a better understanding of the nature of our search results. In particular, it is not clear if an [O/H] model without scatter that fits the data gives a measure of the mean or median metallicity in the IGM.

A model with lognormal metallicity scatter, i.e Gaussian scatter with variance $\sigma_s^2$ in ${\rm log} Z = \rm{[O/H]}$, has
\begin{eqnarray}
{\rm log} Z_{med} & = & median({\rm log} Z) =  \langle {\rm log Z}\rangle  \nonumber \\
& = & {\rm log} \langle Z \rangle -\frac{{\rm ln} 10}{2}\sigma_s^2,
\label{medianmean}
\end{eqnarray}
where ${\rm log} Z_{med}$ and $\langle Z \rangle$ are the median and mean metallicity, respectively, and we adopt $\sigma_s^2=0.5$ (\citealt{s03}, eq. 7).
The lognormal scatter is added on scales of $1.2 h^{-1} {\rm Mpc}$ comoving in regions where $-0.5<{\rm log}(\rho/{\bar \rho})<2$. As equation \ref{medianmean} shows, lognormal models with different scatter, $\sigma_s$, have the same $\langle {\rm log} Z \rangle$ and the same $Z_{med}$ but {\it different} $\langle Z \rangle$.

\begin{figure}
\centering
\mbox{
\includegraphics[angle=90,width=0.97\columnwidth]{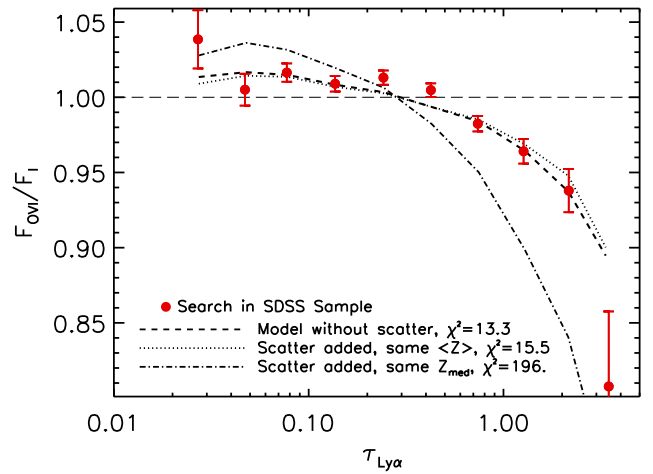}}
\caption{The impact of scatter in our models and its consequences for the nature of the LCP search. The red filled circles are the LCP search as in Figure \ref{searchandnull}. The dashed line is the best fitting model with constant [O/H] and no scatter as in Figure \ref{plotsimav}. We have added lognormal scatter to this model 
while keeping either the mean O/H fixed ({\it dotted} line)
or the median O/H fixed ({\it dot-dashed} line). 
The agreement of the former but not the latter indicates that the scatter-free model fit yields a robust estimate of the mean O/H, not the median O/H (or the mean [O/H]).
Note that the $\tau_{\rm Ly\alpha} \approx 3.3$ point is not included in the fits.
}
\label{plotscatter}
\end{figure}

In order to test whether the LCP search for \OVI\, is sensitive to the mean or median oxygen abundance we explore both interpretations. Figure \ref{plotscatter} compares our best fitting constant [O/H] model with no scatter (as shown in Figure \ref{plotsimav}) to two models with scatter, one that has the same $\langle Z \rangle$ and one that has the same $Z_{med}$.
The good agreement in the former case indicates that our search for \OVI\ is sensitive to the $\langle Z \rangle $ or the mean O/H. Furthermore, we have tested this hypothesis using the A08 density-dependent fit to \OVI\ seen in high resolution spectra.  They report a model fit 
\begin{eqnarray}
\rm{[O/H]} & = & \rm{[O/C]}+\rm{[C/H]} \nonumber \\
& = & {-2.81}^{+0.15}_{-0.14}+0.08^{+0.09}_{-0.06}(z-3) \nonumber \\
& & +0.65^{+0.1}_{-0.14}({\rm log}[\rho/{\bar \rho}]-0.5)
\end{eqnarray}
that is derived from \OVI\ correlated with \CIV\ and \CIV\ correlated with \HI. This model is a three parameter fit to the median [O/H]. Since it is fit to independent data we treat it as having nine degrees of freedom here (equal to the number of data points). When we reproduce this model in our simulations including scatter and perform the LCP search, it provides a reasonable fit to the SDSS data (about $2\sigma$). However, where we have simply added metals to our simulation at this median level without scatter our agreement is poor ({\it reduced} $\chi^2=6$). When we use the correction factor in equation \ref{medianmean} and add metals at the mean level instead, we obtain a reasonable fit once more (at about $1.5\sigma$ confidence).

\section{Measuring the Mean Oxygen Abundance and the Cosmic Oxygen Density}
\label{measure}
Sections \ref{comparison} and \ref{scatter} present a variety of models that are consistent with our LCP search results, despite differences in density dependence and scatter. We can ask whether these models have a common feature of oxygen abundance that is well determined by our data.  
To this end we have calculated the volume-weighted and mass-weighted means of O/H for all models that fit the data: those within $1.5\sigma$ and $2\sigma$ contours, along with the A08 model.  Specifically, we compute the means by summing O/H (not [O/H]) over all pixels in the simulated spectra with mass density in the range $-0.5<{\rm log}[\rho/\bar{\rho}]<2$, either weighting each pixel equally (volume-weighted) or weighting in proportion to mass density.

The mass-weighted mean oxygen abundance, $\mw$, for the `ridge line' of best fitting models at every $(\rho/\bar{\rho})_{\rm cut}$ is shown in Table \ref{meanoonh}. We list the models without scatter, but models with scatter yield the same results since (as shown in Figure \ref{plotscatter}) they must have the same mean oxygen abundance to fit the data. Taking models within the 1.5 and 2$\sigma$ contours (rather than just this ridge line) provides a measure of both uncertainty in density dependence and observational uncertainty.

While the volume-weighted abundances span 0.4-1.3 dex in [O/H], the successful models all have mass-weighted means in the range $-2.51<\mw< -2.39$ ($1.5\sigma$) and $-2.56<\mw< -2.10$ ($2\sigma$).
We conclude that the LCP measurements constrain the mass-weighted mean oxygen abundance of the IGM to be $\mw=-2.45\pm0.06$ ($1.5\sigma$) and $\mw=-2.33\pm0.23$ ($2\sigma$). 
We can compare this result directly with the mass-weighted mean in A08 of $\mw=-2.14\pm0.14$ ($1\sigma$), which is marginally consistent with our results.
The volume-weighted mean oxygen abundance is more weakly constrained to  $\vw=-3.06\pm0.22$ ($1.5\sigma$) and $\vw=-2.74\pm0.64$ ($2\sigma$). These error estimates do not include systematic uncertainties associated with the shape and intensity of the ionizing background (see \S\ref{ioncorr}) or uncertainties in the solar oxygen abundance. Using \citet{c08} values for the solar abundance of oxygen would give a mass-weighted $\mw=-2.34\pm0.06$ ($1.5\sigma$) and $\mw=-2.22\pm0.23$ ($2\sigma$).

With this measure of \mw, and extrapolating to all densities, we can calculate the cosmic oxygen density using
\begin{equation}
\Omega_{\rm Oxy, IGM}\approx Y_{\rm H} A_{\rm O} \Omega_b 10^{\mw}({\rm O/H})_\odot,
\end{equation}
where $Y_{\rm H}=0.76$ is the mass fraction of hydrogen and $A_{\rm O}=16$ is the atomic mass of oxygen. We obtain a value of $\Omega_{\rm Oxy, IGM}=1.4(\pm0.2)\times 10^{-6}$ ($1.5\sigma$). This would be $\sim5\%$ smaller if we limited ourselves to the density range  $-0.5<{\rm log}[\rho/\bar{\rho}]<2$ (the range for which the mass-weighted mean was calculated). This shows that the calculation is not sensitive to the details of the density range integrated over, but this is not a statement about the metal enrichment of systems beyond our constraining density range. In particular, the population of strong absorbers, such as those resulting in the $\tau_{\rm Ly\alpha}\approx3.3$ point of our LCP, are better dealt with in directed searches \citep{f09a,f09b}.

If we follow the procedure of \citet{b07} (and A08) and take $\Omega_{\rm Z, IGM} \approx \Omega_{\rm Oxy, IGM}/0.6$, then we obtain a total cosmic metal density of the IGM, $\Omega_{\rm Z, IGM}\approx 2.3\times 10^{-6}$. This is $\approx 16\%$ of their metal budget of $\Omega_{\rm Z}=1.5 \times 10^{-5}$ at $z=3$ (from their figure 1 and based on integrated cosmic star formation with standard assumptions about about stellar IMF and yields). This metal content is over half that found in galaxies at $z=2.5$ (30\%) while sub-DLAs may contribute between 2 and 17\% \citep{b07}. Summing these values would account for over half of the metals produced by star formation. The remainder could plausibly reside in intermediate densities (e.g. strong \OVI\ absorbers, \citealt{f09b}) or in a hidden warm-hot phase, but uncertainties in the calculated metal production and the observational estimates preclude strong statements about ``missing'' metals.

\begin{table}
\centering
\caption{Mass-weighted oxygen abundance in some best fitting models}
\label{meanoonh}
\begin{tabular}{l c c} 
\hline\hline
Model ([O/H], $(\rho/\bar{\rho})_{\rm cut}$) & \mw & Goodness-of-fit\\
\hline
-2.15, 0		& 	-2.15		&	1.8$\sigma$\\
-2.10, 1		&	-2.25		& 	1.8$\sigma$\\
-2.00, 2		 &    	-2.39		&	1.5$\sigma$\\
-1.80, 3		   &  	-2.41		&	1.3$\sigma$\\
-1.65, 4		&     -2.45		&	1.2$\sigma$\\
-1.45, 5		  &   	-2.43		&	1.3$\sigma$\\
-1.80, 6		   &  	-2.43		&	1.4$\sigma$\\
-1.15, 7		  &   	-2.43		&	1.7$\sigma$\\
-0.95, 8.		 &	-2.36		&	1.8$\sigma$\\
\hline 
A08 			&	-2.36		&     2$\sigma$\\
\hline 
\multicolumn{3}{l}{}                                             \\       
\end{tabular}
\tablecomments{For models along the ridge-line of Figure \ref{chicontour}, with enrichment density cutoffs but no scatter in metallicity at fixed density, we list the [O/H] corresponding to the mass-weighted mean oxygen abundance ratio $\langle {\rm O/H} \rangle$ in column 2 and the acceptability of the $\chi^2$ fit to the data in column 3. }
\end{table}

\section{DISCUSSION}

\subsection{Ionization Corrections}
\label{ioncorr}

We have obtained the most precise constraints to date on the mass-weighted mean O/H and $\Omega_{\rm Oxy}$, but our analysis rests on the assumption that the detected \OVI\ is mostly ionized by a UV background with the spectral shape of the HM01 QG model. The use of this UV background model is motivated by the work of S03 and A08. They also use two other background models: one resulting from quasars only (``Q") and a softened version of the QG model (``QGS") with a lower flux above 4 Ryd. They find that those two models are inconsistent with their relative abundance measures of [O/C] and [O/Si], while they obtain a good fit using the QG model. It is not clear to what extent small variations in the QG model can be tolerated, and what impact these would have on the quoted results. This is clearly an area that requires further investigation.

Our simulations assume a power-law equation of state and so under-produce gas at $T>10^5 {\rm K}$, which is dominated by collisional ionization. This choice is justified in light of studies by \citet{csk02}, \citet{b02} and \citet{ssr04} who show that \OVI\ lines are typically too narrow to be in this phase. It is also notable that \OVI\ which arises from collisional ionization also tends to be strong due to the high \OVI\ fraction in this regime, while the  \OVI\ absorption probed here is relatively weak (even compared to the studies listed above).

A08 find a substantial increase in [O/H], for a fixed [O/C] ratio, for systems that have $\rho/{\bar \rho}>30$. Lower densities are well described by photoionization and a fixed [O/C] ratio in their models. They conclude that $\rho/{\bar \rho}>30$ regions are in fact collisionally ionized (while systems with $\rho/{\bar \rho}<30$ are photoionized). \citet{ssr04} also find such a change in [O/C] and infer a change in ionization mechanism at higher densities. We find indications of the same effect for our measure at $\tau_{{\rm Ly}\alpha}\approx 3.3$, which shows \OVI\ absorption at a level far higher than for lower $\tau_{{\rm Ly}\alpha}$. We have omitted this data point for our photoionization dominated analysis. We have only one point showing this effect, since we do not use higher order Lyman lines in our analysis and are therefore unable to cleanly isolate the highest density regions.  We find a good fit at all detectable \lya\ absorption levels below $\tau_{{\rm Ly}\alpha}\approx 3.3$ using a constant [O/H] and assuming a photoionized medium.

\begin{figure}
\centering
\mbox{
\includegraphics[angle=90,width=0.98\columnwidth]{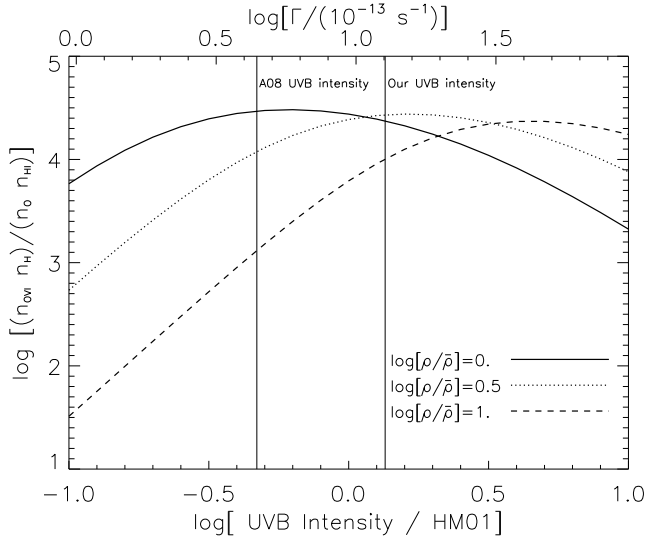}}
\caption{The ionization correction parameter $(n_{\rm OVI}n_{\rm H})/ (n_{\rm O}n_{\rm HI})$ in terms of the UV background intensity (normalized to the HM01 level) and ionization rate at $z=3$. This is calculated for 3 densities  that span our range of sensitivity: mean density ({\it solid} line), ${\rm log}[\rho/\bar{\rho}]=0.5$ ({\it dotted} line), and ${\rm log} [\rho/\bar{\rho}]=1$ ({\it dashed} line). The thin vertical lines show the UVB intensity used in A08 and here.}
\label{intencompare}
\end{figure}

One should note that A08 normalise their UV background to a level that differs from ours. This leads to an adjustment on the ionization correction. The ratio of optical depths is related to the metallicity by
\begin{equation}
{\rm log}\Big[\frac {\tau_{\rm OVI}}{\tau_{\rm Ly\alpha}}\Big]={\rm log}\Big[\frac{n_{\rm OVI}}{n_{\rm O}}\frac{n_{\rm H}}{n_{\rm HI}}\Big]+{\rm [O/H] +log (O/H)_\odot }+S,
\label{ioncorreq}
\end{equation}
where $S={\rm log}[(f\lambda)_{\rm OVI}/(f\lambda)_{\rm Ly\alpha}]$. The first term on the right hand side of this equation is the ionization correction term, and this is plotted against UV background intensity (normalized to the HM01 level) and the ionization rate in Figure \ref{intencompare}. This was calculated assuming the equation of state in equation \ref{equofstate} and $z=3$ for three overdensities: ${\rm log}[\rho/\bar{\rho}]=0$, 0.5 and 1 that span our sensitivity range. Our chosen UV background intensity of $1.35\times{\rm HM01}$ ($\Gamma=13\ee{-13} {\rm s^{-1}}$) is shown along with the A08 value of $0.48\times{\rm HM01}$ ($\Gamma=4.6\ee{-13} {\rm s^{-1}}$). For mean density the ionization correction is largely unchanged between our simulations and those of A08, however, our models would result in systematically lower [O/H] resulting from oxygen in systems of ${\rm log}[\rho/\bar{\rho}]>0.5$ by between 0.3 and 1.1 dex. This might go some way to explaining the 0.2 dex lower value we find for [O/H]. Another source of potential discrepancy is differences in the density distributions between the studies. Despite these issues, we are in broad agreement with a metal budget fraction of $\sim 15 - 35\%$ in A08 for $z=2-3$.

The success of the LCP search method for detection of \OVI\ in the SDSS spectra may indicate that it is appealing for other searches for low level absorption in the IGM, where large numbers of moderate resolution spectra with moderate signal-to-noise and a substantial uncertainty in the continuum are available. There are a number of other species that may be detectable in the SDSS sample with a similar approach. Correlation searches in spectra from other instruments such as FUSE and STIS, or even HIRES and UVES, may benefit from this kind of analysis, which is resistant to errors in the continuum. Modulation of the number of local pixels used may also be a useful approach to measuring clustering of absorbers.

\subsection{Continuum Correction Uncertainty}

We systematically underestimate our continuum fitting such that the continua must be corrected up (and the transmitted flux down) by 16\%  to match the mean flux measured in studies using high resolution spectra. The uncertainty in those measurements may lead to systematic errors in our measurements. We use $\langle F \rangle = 0.68 \pm 0.04$ at$z=3$ from \citet{m00} and those $1\sigma$ error estimates correspond to allowed corrections of 11-25\%. This would lead to a 15\% offset in  $\langle \tau_{{\rm Ly}\alpha} \rangle$, which would be corrected by a 15\% offset in the recombination rate. This change in recombination rate results in change to the \HI\ fraction which is independent of density for the regime probed here, leaving the \lya\ forest largely unchanged. There is a small distortion due to the fact that an error which is a multiple in flux would be corrected by a factor which is a multiple in optical depth, but
crucially most of the signal in the LCP search comes from higher optical depths for which the correction is even smaller. Also large errors of this sort would be evident in the comparison between the PDF of the flux in the simulations and the data in Figure \ref{pdfcompare}.

A systematic error in the chosen $\Gamma$ (based on an error in the mean flux
in the literature) would also impact upon the \OVI\ fraction.  This error corresponds to half a minor tick interval in the upper horizontal axis of Figure \ref{intencompare} and so ionization correction parameter would be largely unchanged over the density range of interest.
One can conclude that in the context of (related) uncertainties in the shape 
and amplitude of the UV background and so $\Gamma$, these issues are a
sub-dominant systematic uncertainty.

\subsection{Other Searches for \OVI\ in the IGM}

\citet{fbp08} argue that the lack of velocity dependence in the column densities ($N$) of their proximate ($\delta v< 8000\ \kms$ from $z_{\rm QSO}$) sample indicates that their 26 weak \OVI\ absorbers must be collisionally ionized. The majority of these absorbers likely correspond to systems of $\rho/{\bar \rho}>10$ based on their associated \HI\ column densities and using the $N-\rho/{\bar \rho}$ relation from \citet{s01}. Hence they are sampling predominantly stronger absorbers than those considered here, so this work is not at odds with our choice of a photoionization model for our calculation of [O/H]. 

Our results appear consistent with other comparable measures of [O/H] in the IGM, given the differences of method and thus the systems sampled.
\citet{csk02} also use high resolution spectra to search for \OVI, but unlike \citet{a08} they use Voigt profile fitting of detected lines. They argue that these lines at $z=2$ are photoionized (and so are comparable with those we find), with abundances $10^{-3}-10^{-2}$ of solar.

\citet{t02} use stacked \OV\ in FUSE spectra to determine [O/H] and obtain a range of -2.2 to -1.3. Their higher value is partly caused by a combination of their use of a ``Q'' UV background model and by their lower redshift range ($1.6<z<2.9$). The main cause for the discrepancy is likely their use of the 78th percentile of their stack (in order to maximize signal-to-noise), while we use the 50th. \citet{a08} argue that this difference likely corresponds to a 0.5 dex adjustment, bringing the results of \citet{t02} broadly into line with ours.

 \citet{bh05} search for \OVI\ lines at somewhat lower redshift range ($2<z<2.6$) than ours. They divide these lines as `metal-poor' and `metal-rich' . This differs greatly from our analysis and so is difficult to compare.  In particular they find the [O/H] of systems with detected \OVI\ absorption while we calculate the average metallicity over all \lya\ absorption. 
Their `metal-poor' sample shows $-3<{\rm[O/H]}<-1$ and a cosmic oxygen density $\Omega_{\rm Oxy}\approx 2.3 (\pm^{2.1}_{0.6})\ee{-6}$ using a photoionization model similar to the ``Q'' model described above, which is consistent with our results, but the numbers are not directly comparable. 

\citet{f09a} have searched the same SDSS dataset with the goal of identifying strong \OVI\ absorbers, requiring additionally associated \lya\ and \lyb\ absorption. The approach used is complementary to this work, as they can only retrieve strong absorbers that are typically saturated in both these Lyman lines. Hence, these absorbers are a different population from the ones probed here with higher density and (perhaps) collisional ionization.
\citet{f09b} provide a measure of the cosmic oxygen density contributed by these systems of $\Omega_{\rm O VI} \geq 1.9 \times 10^{-8} h^{-1}$ (which corresponds to a firm lower limit of $\Omega_{\rm Oxy} \geq 1 \times 10^{-7} h^{-1}$ assuming the maximum \OVI\ ionization fraction of 20\%) at absorber redshifts $2.8 \leq z_{abs} \leq 3.2$. 

Our measure of the fraction of the estimated metal budget detected at $z=3$ is in line with values calculated by \citet{b07} using results from various papers and various metals. They infer a `forest' contribution of $<15\%$ at $z=2$. One study they quote is particularly notable: \citet{ssr04} find $\Omega_{\rm Oxy}\approx 2 \ee{-6}$ using a  `hard' (`Q' like) UV background.  \citet{b07} reanalyze  these results in the context of a `softer' (`QG' like) UV background model and find  $\Omega_{\rm Oxy}\approx 5.0 \ee{-6}$ at $z=2.5$, and so a fraction of the metal budget of $\approx 30\%$. This measure is consistent with our findings given redshift evolution and systematic uncertainties associated with search methods and ionization corrections. \citet{ssr04} find a mean $\rm [O/H]=-2.85$ for their lines (which is effectively a volume-weighted mean) for the 70\% of systems in which they detected oxygen, which is also consistent with our findings (setting the other 30\% to zero metallicity would result in a volume-weighted mean of $\vw=-3.00$).

\section{CONCLUSIONS}
\label{conclusions}

We have developed a new method to detect \OVI\ absorption in low-density regions with a data set 
that might initially seem ill suited to this purpose, the moderate resolution ($R=1800$), moderate signal-to-noise ratio SDSS spectra of high-redshift QSOs.
Not only have we successfully detected \OVI\ with high significance (the null case is ruled out with  {\it reduced} $\chi^2=10$), but we have placed tighter limits on the mass-weighted mean oxygen abundance of low-density regions than ever before. The success of this method hinges on the sheer size of the SDSS sample of spectra and the use of local, uncorrelated pixels near the \OVI\ absorber as a way of characterizing both the continuum fitting error and the degree of contaminating absorption.  This locally calibrated pixel (LCP) search for \OVI\
seen in SDSS DR3 is the main result of the paper and the high precision of this
new tracer for metals in the IGM is clear. Modeling has been necessary to interpret these results and conclusions about oxygen abundance the lognormal random fields have been used to produce more than 4.5 million synthetic spectra.

Our search at $2.7<z<3.2$ is not sensitive to the density dependence of the [O/H] because of spectral resolution constraints, but using detailed modeling we are able to measure the oxygen abundance IGM densities of $1 \lesssim \rho/\bar{\rho} \lesssim 9$ and can conclude the following:

\begin{itemize}

\item A model with constant $\rm{[O/H]}=-2.15_{-0.09}^{+0.07}$ provides an adequate fit to the data (at the 2$\sigma$ level), but a fit can also be obtained by removing metals from lower density regions and elevating the metallicity in regions enriched. 

\item The addition of lognormal metallicity scatter leaves the quality of fit of these models largely unchanged, where the scatter free model is treated as a measure of the mean O/H (or $\langle  Z \rangle$)  and {\it not} the mean [O/H] (or $\langle log  Z \rangle$).

\item The mass-weighted mean oxygen abundance is nearly constant among our viable models. As a result we have been able to place the tightest constraints thus far on this quantity, with $\mw=-2.45\pm0.06$  (at $1.5\sigma$ confidence) computed over the density range $-0.5<{\rm log}[\rho/\bar{\rho}]<2$. This value is based on a \citet{gns96} solar abundance of oxygen. A \citet{c08} solar abundance would give a abundance $\mw=-2.34\pm0.06$.

\item The models that fit the data have volume-weighted mean, $\vw=-3.01\pm0.33$ (at $1.5\sigma$ confidence).

\item We calculate the cosmic density of oxygen, $\Omega_{\rm Oxy, IGM}=1.4(\pm0.2)\times 10^{-6}$. In the context of the work by \citet{b07} this constitutes a total metal contribution of $\Omega_{\rm Z, IGM}\approx 2.3\times 10^{-6}$, which is $\approx 16\%$ of their estimated metal budget at  at $z=3$.
\end{itemize}

This novel LCP search method need not be limited to searches for metals in the IGM. It could be adapted to spectral searches wherever bulk, low-level correlated absorption is expected and continuum errors and contaminating absorbers are the limiting factor. Little prior knowledge of these errors and uncertainties is required, only that they are uncorrelated.

\acknowledgments

We thank Arlin Crotts, J. Xavier Prochaska and Romeel Dav\'e for their useful comments.

Funding for the SDSS and SDSS-II has been provided by the Alfred P. Sloan Foundation, the Participating Institutions, the National Science Foundation, the U.S. Department of Energy, the National Aeronautics and Space Administration, the Japanese Monbukagakusho, the Max Planck Society, and the Higher Education Funding Council for England. The SDSS Web Site is \url{http://www.sdss.org/}.

The SDSS is managed by the Astrophysical Research Consortium for the Participating Institutions. The Participating Institutions are the American Museum of Natural History, Astrophysical Institute Potsdam, University of Basel, University of Cambridge, Case Western Reserve University, University of Chicago, Drexel University, Fermilab, the Institute for Advanced Study, the Japan Participation Group, Johns Hopkins University, the Joint Institute for Nuclear Astrophysics, the Kavli Institute for Particle Astrophysics and Cosmology, the Korean Scientist Group, the Chinese Academy of Sciences (LAMOST), Los Alamos National Laboratory, the Max-Planck-Institute for Astronomy (MPIA), the Max-Planck-Institute for Astrophysics (MPA), New Mexico State University, Ohio State University, University of Pittsburgh, University of Portsmouth, Princeton University, the United States Naval Observatory, and the University of Washington.

%

%

%


\begin{thebibliography}{27}

	
	
	
\bibitem[Abazajian et al.(2005)]{aetal05}
Abazajian, K. et al. 2005, \aj, 129, 1755

\bibitem[{Adelberger} {et~al.}(2003)]{a03}
{Adelberger}, K.~L., {Steidel}, C.~C., {Shapley}, A.~E., \& {Pettini}, M. 2003,
  \apj, 584, 45

\bibitem[{Adelberger} {et~al.}(2005)]{a05}
{Adelberger}, K.~L., {Shapley}, A.~E., {Steidel}, C.~C., {Pettini},
M., Erb, D.~K., \& Reddy, N.~A. 2005,
  \apj, 629, 636

\bibitem[{{Aguirre}  et~al.}(2002)]{a02}
{Aguirre} A.,  {Schaye} J., \&  {Theuns} T.,  2002, ApJ, 576, 1

\bibitem[{{Aguirre}  et~al.}(2008)]{a08}
 {Aguirre} A., {Dow-Hygelund} C., {Schaye} J., \& {Theuns}, T.
2008, \apj, 689, 851 (A08)

\bibitem[{{Asplund}  et~al.}(2005)]{ags05}
Asplund, M., Grevesse, N., \& Sauval, A.~J. 2005 ASP Conf. Ser., ed Barnes, III, T.~G. \& Bash, F.~N.

\bibitem[Basu \& Antia(2004)]{ba04}
{Basu}, S. and {Antia}, H.~M. 2004, \apjl, 606, L85

\bibitem[{{Bergeron} et~al.}(2002)]{b02}
{ Bergeron} J., {Aracil} B., {Petitjean} P., \& {Pichon}, C.
2002, A\&A, 396, L11

\bibitem[{Bi} {et~al.}(1992)]{bbc92}
{Bi} H.~G., {Boerner} G., \& {Chu} Y., 1992, A\&A, 266, 1

\bibitem[{Bi} \& {Davidsen}(1997)]{bd97}
{Bi} H., \& {Davidsen} A.~F., 1997, ApJ, 479, 523

\bibitem[Bergeron \& Herbert-Fort(2005)]{bh05}
Bergeron, J., \& Herbert-Fort, S. 2005, (astro-ph/0506700)

\bibitem[{{Bouch\'e} et~al.}(2007)]{b07}
{ Bouch\'e} N., {Lehnert} M., {Aguirre} A., {P\'eroux}, C., \& Bergeron J. 2007, 378, 525

\bibitem[Carswell et~al.(2002)]{csk02}
{{Carswell}, B., {Schaye}, J., \& {Kim}, T.-S.},
 2002, \apj, 578, 43

\bibitem[Caffau et~al.(2008)]{c08}
Caffau, E., Ludwig, H.-G., Steffen, M., Ayres, T.~R., 
Bonifacio , P., Cayrel, R., Freytag, B., \& Plez, B. 2008, A\&A, 488, 1031

\bibitem[Cowie et~al. (1995)]{c95}
Cowie, L.~L., \& Songaila, A., Kim, T.-S. , \& Hu, E.,~M. 1995, \aj, 109, 1522


\bibitem[Cowie \& Songaila(1998)]{cs98}
Cowie, L.~L., \& Songaila, A. 1998, \nat, 394, 44

\bibitem[Croft et~al.(1998)]{c98}
{{Croft}, R.~A.~C., {Weinberg}, D.~H., {Katz}, N., \& {Hernquist}, L.} 1998, \apj, 495, 44


\bibitem[Dav\'e et~al.(1998)]{d98}
{{Dav{\'e}}, R., {Hellsten}, U., {Hernquist}, L., {Katz}, N.,\& 
	{Weinberg}, D.~H.} 1998, \apj, 509, 661


\bibitem[Delahaye \& Pinsonneault(2006)]{dp06}
{Delahaye}, F., \& {Pinsonneault}, M.~H. 2006, \apj, 649, 529

\bibitem[{Efstathiou} {et~al.}(1992)]{ebw92}
{Efstathiou} G., {Bond} J.~R., \& {White} S.~D.~M., 1992, \mnras, 258, 1P

\bibitem[Ellison et~al.(2000)]{e00} 
Ellison, S.~L., Songaila, A., Schaye, J., \& Pettini, M. 2000, \apj, 120, 1175

\bibitem[Ferland et~al(1998)]{f98}
Ferland, G.~J., Korista, K.~T., Verner, D.~A., Ferguson, J.~W., Kingdon, J.~B., \& Verner,  E.M. 1998, PASP, 110, 761

\bibitem[Fox et~al.(2008)]{fbp08}
Fox, A., Bergeron, J., \& Petitjean, P. 2008, \mnras, 388,1557

\bibitem[Frank et~al.(2009a)]{f09a}
Frank, S., Mathur, S., York, D., \& Pieri M. M. 2009, submitted to AJ, (astro-ph/0707.1700)

\bibitem[Frank et~al.(2009b)]{f09b}
Frank, S., Mathur, S., York, D. \& Pieri M. M. 2009, submitted to AJ, (astro-ph/0806.3071)

\bibitem[Gibson et~al.(2009)]{g09}
{{Gibson}, R.~R., {Jiang}, L., {Brandt}, W.~N., {Hall}, P.~B., 
	{Shen}, Y., {Wu}, J., {Anderson}, S.~F., {Schneider}, D.~P., 
	{Vanden Berk}, D., {Gallagher}, S.~C., {Fan}, X.,\& {York}, D.~G.}
	2009, \apj, 692, 758

\bibitem[Grevesse et~al.(1996)]{gns96}
Grevesse, N., Noels, A., \& Sauval, A.~J. 1996, in ASP Conf. Ser., ed. S. S. Holt and G. Sonneborn , 99, 117

\bibitem[Haardt \& Madau(2001)]{hm01}
{{Haardt}, F., \& {Madau}, P.} 2001, in Clusters of Galaxies and the High Redshift Universe Observed in X-rays, ed. {Neumann}, D.~M. \& {Tran}, J.~T.~V., (astro-ph/0106018) (HM01)

\bibitem[{{Hellsten} et~al.}(1998)]{h98}
{Hellsten} U.,  {Hernquist} L.,  {Katz} N.,\& {Weinberg} D.~H.,  1998, \apj,
  499, 172
  
\bibitem[Hui \& Gnedin(1997)]{hg97}
Hui, L., \&  Gnedin 1997, \mnras, 292, 27
  

\bibitem[{{Kaiser} \& {Peacock}(1991)}]{kp91}
{Kaiser} N., \& {Peacock} J.~A. 1991, ApJ, 379, 482

\bibitem[{Kim} {et~al.}(2007)]{k07}
{{Kim}, T.-S., {Bolton}, J.~S., {Viel}, M., {Haehnelt}, M.~G., \& 
	{Carswell}, R.~F.} 2007, \mnras, 382,1674

\bibitem[{Komatsu} {et~al.}(2008)]{k08}
{{Komatsu}, E., {Dunkley}, J., {Nolta}, M.~R., {Bennett}, C.~L., 
	{Gold}, B., {Hinshaw}, G., {Jarosik}, N., {Larson}, D., 
	{Limon}, M., {Page}, L., {Spergel}, D.~N., {Halpern}, M., 
	{Hill}, R.~S., {Kogut}, A., {Meyer}, S.~S., {Tucker}, G.~S., 
	{Weiland}, J.~L., {Wollack}, E.\& {Wright}, E.~L.}
	2008, submitted to ApJS, (astro-ph/0803.0547)
	
\bibitem[Mac Low \& Ferrara(1999)]{mf99} Mac Low, M.-M., \&
    Ferrara, A. 1999, \apj, 513, 142

\bibitem[Madau, Ferrara, \& Rees(2001)]{mfr01}
    Madau, P., Ferrara, A., \& Rees, M. J. 2001, \apj, 555, 92

\bibitem[Meyer \& York(1987)]{my87}
Meyer, D.~M., \& York, D.~G. 1987, \apj, 315, L5


\bibitem[McDonald et al.(2006)]{m00}
{{McDonald}, P., {Miralda-Escud{\'e}}, J., {Rauch}, M., 
{Sargent}, W.~L.~W. , {Barlow}, T.~A., {Cen}, R.,\& {Ostriker}, J.~P.} 2000, \apj, 543, 1


\bibitem[McDonald et al.(2006)]{m06}
{{McDonald}, P., {Seljak}, U., {Burles}, S., {Schlegel}, D.~J., 
	{Weinberg}, D.~H., {Cen}, R., {Shih}, D., {Schaye}, J., 
	{Schneider}, D.~P., {Bahcall}, N.~A., {Briggs}, J.~W., 
	{Brinkmann}, J., {Brunner}, R.~J, {Fukugita}, M., 
	{Gunn}, J.~E., {Ivezi{\'c}}, {\v Z}., {Kent}, S., {Lupton}, R.~H.,\& 
	{Vanden Berk}, D.~E.} 2006, \apjs, 163, 80
	
\bibitem[Oppenheimer \& Dav\'e(2006)]{od06}
    Oppenheimer, B.~D., \& Dav\'e, R. 2006, \mnras, 373, 1265


\bibitem[Pettini et al.(2001)]{p01}
Pettini, M., Shapley, A.~E., Steidel, C.~C., Cuby, J., Dickinson, M., Moorwood, A.~F.~M., Adelberger, K.~L., \& Giavalisco, M. 2001, \apj, 554, 981

 \bibitem[{{Pieri} \& {Haehnelt}(2004)}]{ph04}
{Pieri}, M.~M., \& {Haehnelt}, M.~G. 2004, \mnras, 347, 985

\bibitem[Pieri at~al.(2006)]{psa06}
{Pieri}, M.~M., Schaye, J., \& Aguirre, A.,\apj, 638, 45

 \bibitem[Pieri, Martel, \& Grenon(2007)]{pmg07}
    Pieri, M.~M., Martel, H., \& Grenon, C. 2007, \apj, 658, 36

 \bibitem[Pieri \& Martel(2007)]{pm07}
    Pieri, M.~M. \& Martel, H. 2007, \apjl, 662, 36, L7
        

\bibitem[{{Rauch} et~al.}(1997)]{rhs97}
{Rauch} M.,  {Haehnelt} M.~G., \&  {Steinmetz} M.,  1997, \apj, 481, 601

\bibitem[{{Rauch} et~al.}(1997)]{r97}
{{Rauch}, M. and {Miralda-Escude}, J., {Sargent}, W.~L.~W., 
	{Barlow}, T.~A., {Weinberg}, D.~H., {Hernquist}, L., 
	{Katz}, N., {Cen}, R., \& {Ostriker}, J.~P.} 1997, \apj, 489, 7
    
\bibitem[Scannapieco et al.(2006)]{s06} 
Scannapieco, E., 
Pichon, C., Aracil, B., Petitjean, P., Thacker, R.~J., Pogosyan, D., 
Bergeron, J., \& Couchman, H.~M.~P. 2005, \mnras, 365, 615

\bibitem[{Schaye} {et~al.}(2003)]{s03}
    Schaye, J., Aguirre, A., Kim, T.-S., Theuns, T., Rauch, M., \&
    Sargent, W.~L.~W. 2003, \apj, 596, 768 (S03)

\bibitem[Schaye(2001)]{s01}
	Schaye, J. 2001, \apj, 559, 507

\bibitem[{Simcoe} et~al.(2002)]{ssr02}
{{Simcoe} R.~A, {Sargent} W.~L.~W., \& {Rauch}, M.}
2002, \apj, 578, 737
	

\bibitem[{Simcoe} et~al.(2004)]{ssr04}
{{Simcoe} R.~A, {Sargent} W.~L.~W., \& {Rauch}, M.}
2004, \apj, 606, 92

\bibitem[Songaila \& Cowie(1996)]{sc96}
Songaila, A., \& Cowie, L.~L. 1996, \aj, 112, 335

\bibitem[Telfer et al.(2002)]{t02}
Telfer, R.~T., Kriss, G., Zheng, W., Davidsen, A.~F., \& Tytler, D. 2002, \apj, 579, 500

\bibitem[Viel et al.(2002)]{v02}
 {{Viel}, M., {Matarrese}, S., {Mo}, H.~J., {Theuns}, T., \&
	{Haehnelt}, M.~G.} 2002, /mnras, 336, 685

\bibitem[Weinberg et al.(1998)]{w98}
Weinberg, D.~H., Katz, N., \& Hernquist, L. 1998, ASP Conf. Ser., ed Woodward, C.~E., Shull, J.~M., \& Thronson, H.~A.

\bibitem[Wild et al.(2008)]{w08}
Wild, V., Kauffmann, G., White, S., York, D., Lehnert, M., Heckman, T.,Hall, P.~B, 
Khare, P., Lundgren, B. Schneider, D.~P., \& Vanden Berk, D.
2008, \mnras, 338, 227


\bibitem[Yip et~al.(2004)]{2004AJ....128.2603Y}
{{Yip}, C.~W., {Connolly}, A.~J., {Vanden Berk}, D.~E., 
	{Ma}, Z., {Frieman}, J.~A., {SubbaRao}, M., {Szalay}, A.~S., 
	{Richards}, G.~T., {Hall}, P.~B., {Schneider}, D.~P., 
	{Hopkins}, A.~M., {Trump}, J., \& {Brinkmann}, J.} 2004, \aj, 128, 2603
	
\bibitem[York et~al.(2000)]{y00}
	York, D. G., et al. 2000, \aj, 120, 1579 
	
\bibitem[York et~al.(2006)]{y06}
	{York}, D.~G, {Khare}, P., {Vanden Berk}, D., {Kulkarni}, V.~P., 
	{Crotts}, A.~P.~S., {Lauroesch}, J.~T., {Richards}, G.~T., 
	{Schneider}, D.~P., {Welty}, D.~E., {Alsayyad}, Y., 
	{Kumar}, A., {Lundgren}, B., {Shanidze}, N., {Smith}, T., 
	{Vanlandingham}, J., {Baugher}, B., {Hall}, P.~B., 
	{Jenkins}, E.~B., {Menard}, B., {Rao}, S., {Tumlinson}, J., 
	{Turnshek}, D., {Yip}, C.-W., \& {Brinkmann}, J. 2006, \mnras, 367, 945
	
	

    
\end{thebibliography}
\end{document}